\documentclass{rmaa}

\title{Proper motions of the HH~1 jet}
\author{A. C. Raga
\affil{Instituto de Ciencias Nucleares, UNAM}
B. Reipurth
\affil{Institute for Astronomy, Univ. of Hawaii}
A. Esquivel, A. Castellanos-Ram\'\i rez, P. F. Vel\'azquez, L. Hern\'andez-Mart\'\i nez,
A. Rodr\'\i guez-Gonz\'alez, J. S. Rechy-Garc\'\i a, D. Estrella-Trujillo
\affil{Instituto de Ciencias Nucleares, UNAM}
J. Bally
\affil{CASA, Univ. of Colorado}
D. Gonz\'alez-G\'omez
\affil{DAFM, Univ. de las Am\'ericas, Puebla}
A. Riera
\affil{Universitat Politecnica de Catalunya}
}

\fulladdresses{
\item A. C. Raga \& A. Castellanos-Ram\'\i rez, P. F. Vel\'azquez, L. Hern\'andez-Ram\'\i rez,
A. Rodr\'\i guez-Gonz\'alez, J. S. Rechy-Garc\'\i a, D. Estrella-Trujillo: Instituto de Ciencias Nucleares,
Universidad Nacional Aut\'onoma de M\'exico, Ap. 70-543, 04510 D. F.,
M\'exico
(raga@nucleares.unam.mx)
\item B. Reipurth: Institute for Astronomy, University of Hawaii at Manoa, Hilo, HI 96720, USA
\item J. Bally: Center for Astrophysics and Space Astronomy, University of Colorado, UCB 389,
Boulder, CO 80309, USA
\item D. Gonz\'alez-G\'omez, DAFM, UDLAP, Ex Hda. Sta. Catarina M\'artir, C:P. 72810, Puebla, M\'exico
\item A. Riera: Departament de F\'\i sica i Enginyeria Nuclear, EUETIB, Universitat Polit\'ecnica de Catalunya,
  Comte d'Urgell 187, E-08036 Barcelona, Espa\~na
}

\shortauthor{Raga et al.}
\shorttitle{Proper motions of the HH~1 jet}

\SetVolume{40} \SetFirstPage{001} \SetYear{2016}
\ReceivedDate{\today} 
\AcceptedDate{Year Month Day} 

\resumen{Describimos un nuevo m\'etodo para determinar movimientos propios de objetos extendidos,
  y un c\'odigo que desarrollamos para la aplicaci\'on de este m\'etodo. Aplicamos este
  m\'etodo a un an\'alisis de cuatro \'epocas de im\'agenes del HST de [S~II] del jet de HH~1 (cubriendo
  un per\'\i odo de $\sim 20$ a\~nos).
  Determinamos los movimientos propios de los nudos a lo largo del jet, y
  hacemos una reconstrucci\'on de la historia de la variabilidad de la velocidad de eyecci\'on (suponiendo nudos
  bal\'\i sticos). La reconstrucci\'on muestra una ``aceleraci\'on'' de la velocidad de eyecci\'on de los
  nudos del jet, con velocidades
  mayores a tiempos m\'as recientes. Esta aceleraci\'on resultar\'a en que los nudos que ahora observamos a lo
  largo del jet se junten en $\sim 450$~a\~nos y a una distancia de $\sim 80''$
  de la fuente, cercana a la posici\'on actual de HH~1.}

\abstract{We describe a new method for determining proper motions of extended objects, and
  a pipeline developed for the application of this method. We then apply this method to an
  analysis of four epochs of [S~II] HST images of the HH~1 jet (covering a period of $\sim 20$~yr).
  We determine the proper motions of the knots along the jet, and make a reconstruction of the past
  ejection velocity time-variability (assuming ballistic knot motions). This reconstruction shows an
  ``acceleration'' of the ejection velocities of the jet knots,
  with higher velocities at more recent times. This acceleration
  will result in an eventual merging of the knots in $\sim 450$~yr and at a distance of $\sim 80''$
  from the outflow source, close to the present-day position of HH~1.}


\begin{document}
\maketitle

\section{Introduction}

The HH~1/2 outflow (discovered by Herbig 1951 and Haro 1952)
has played a fundamental role in the study of collimated flows from young stellar
objects (YSOs), and the associated observational and theoretical work has been reviewed by
Raga et al. (2011). This system has two bright ``heads'': HH~1 (to the NW) and HH~2 (to the SE),
centered on the ``VLA~1'' radio continuum source (Pravdo et al. 1985).

The VLA~1 source also has a jet/counterjet system visible at IR wavelengths (Noriega-Crespo \&
Raga 2012) extending out towards HH~1 and 2. Optically, only the slightly blueshifted N
jet (pointing to HH~1) is visible (Bohigas et al. 1985; Strom et al. 1985), as shown in Figure 1. This optical
feature has been called the ``HH~1 jet''. Apart from the
papers mentioned above, a limited number of papers have studied some of
the characteristics of the HH~1 jet:
\begin{itemize}
\item optical images and proper motions: Reipurth et al. (1993), Eisl\"offel et al. (1994), Bally et al. (2002),
\item radio proper motions: Rodr\'\i guez et al. (2000),
\item infrared images: Davis et al. (2000), Reipurth et al. (2000),
\item infrared spectra: Eisl\"offel et al. (2000), Garc\'\i a L\'opez et al. (2008).
\end{itemize}

Some of the most striking characteristics of HH~1/2 are their proper motions (Herbig \&
Jones 1981; Eisl\"offel et al. 1994; Bally et al. 2002; Hartigan et al. 2011) and
time-variability (Herbig 1969, 1973; Raga et al. 1990; Eisl\"offel et al. 1994). The
fact that there are now four epochs of HST images of HH~1/2, covering a time span
of $\sim 20$~yr (Raga et al. 2015a, b, c; 2016a, b, c) has allowed progress on both of these issues.

Raga et al. (2016a, b) have used the 4 epochs of HST images to determine proper motions
of HH~1 and 2, finding a small acceleration for the motion of HH~1 and a small braking for
HH~2 (when comparing their proper motions to the ones of Herbig \& Jones 1981). They also
used the photometrically calibrated HST images (Raga et al. 2016c) to evaluate the recent time-variability
of the emission of HH~1 and 2 (comparing their line fluxes to the ones of Brugel et al. 1981).

For their study of HH~1/2 proper motions, Raga et al. (2016a, b) explored a new method for
determining motions of angularly extended objects, based on a two-step process:
\begin{itemize}
\item convolving the frames of the different epochs with wavelets of chosen widths,
\item spatially fitting the peaks in the (degraded angular resolution) convolved frames.
\end{itemize}

In the present paper, we apply this new method to the four available epochs of HH~1/2 HST [S~II] images,
in order to determine the proper motions and intensity variations of the knots along the HH~1 jet
(which was not studied in the papers of Raga et al. 2016a, b). We also present a detailed description
of the method, and describe a pipeline (written in Python) developed for applying this method to observational
or simulated emission map time-series.

The paper is organized as follows. Section 2 reviews the methods that have been used to measure
proper motions in CCD frames of HH outflows. Section 3 presents the new method for deriving proper motions and
intensities of extended structures, and describes the Python pipeline.
Section 4 describes the proper motions of the knots along the HH~1 jet, and Section 5
the time-variability of the [S~II] emission. Section 6 describes the standard attempts at using
the observed proper motions to reconstruct the history of the time-variability of the ejection and
to predict the future evolution of the ejected material. Finally, the results are summarized in
Section 7.

\section{Proper motions and time-variabilites of HH objects from CCD images}

As far as we are aware, the first attempt at measuring positions and fluxes of condensations
in CCD frames of HH objects was done by Raga et al. (1990), who analyzed H$\alpha$ and
[O~III]~5007 images of HH~1/2. These authors found the then non-trivial result that even
though they had only two stars in their CCD frames (and were therefore only able to
compute a scaling, rotation and translation rather than a ``real''
astrometric calibration of the images) they obtained positions for the HH~1/2 condensations
that coincided with the forward time-projection obtained with the photographic
proper motions of Herbig \& Jones (1981).

Raga et al. (1990) measured the positions (and peak intensities) of the HH~1/2 condensations
by carrying out paraboloidal fits to the emission peaks seen in the images. This kind of
``peak fitting'' procedure (fitting mostly either a paraboloid or a Gaussian) has been extensively
used for obtaining proper motions of HH outflows (see, e.g., Eisl\"offel \& Mundt 1992, 1994;
Eisl\"offel et al. 1994).

Heathcote \& Reipurth (1992) tried a different method to obtain proper motions from
CCD images of HH outflows. In their analysis of images of HH~34, they defined
a box (including the emission of the HH~34 jet) within which they carried out
cross-correlations between pairs of images. This method proved to be a major
improvement in determining proper motions of HH outflows, as instead of relying
on the positions of sometimes ill-defined peaks, the proper motions are determined
with the emission within a spatially more extended box. This process yields
a cross-correlation function with a much better signal-to-noise ratio (compared to
the images themselves), the peak of which can be fitted to a many times surprising
accuracy. This cross-correlation technique has become the standard method for determining proper
motions of HH objects (see, e.g., Curiel et al. 1997; Reipurth et al. 2002; Hartigan et al. 2005;
Anglada et al. 2007).

The main inconvenience of the cross-correlation method is the fact that one
has to choose boxes of arbitrary shapes (mostly square boxes have been used),
sizes and locations so as to include features that one judges to be well defined
``entities'' within the images. This is of course inconvenient in images
with complex structures of different sizes, and also somewhat problematic since
the determined proper motions clearly depend on the ``cross correlation boxes'' that
have been chosen.

In a study of a planetary nebula, Szyskza et al. (2011) used the interesting method
of covering the images with a regular array of cross-correlation boxes (which they
call ``tiles''). The shifts of the peaks of the correlation functions corresponding
to these boxes then give a ``proper motion map'' of the whole field (actually, a low-intensity
cut-off has to be imposed so as not to obtain random motions in boxes with no visible
emission structures). Raga et al. (2012a, 2013), applied this method to HH objects (with
the implementation of the method being presented in detail in the latter paper).

This method of cross-correlation ``tiles'' has the clear advantage that one only needs
to define:
\begin{itemize}
\item a size for the tiles,
\item a ``beginning point'' at which to begin to draw one of the tiles,
\item a ``low intensity cutoff'' necessary for the proper motions to be calculated.
\end{itemize}
These are of course many fewer free parameters than the ones involved in a ``free choice''
cross-correlation box scheme.

However, it is evident that there are complications to this method. Two of these are that:
\begin{itemize}
\item the tiles sometimes include only part of an apparently coherent structure (algorithmical
  efforts to surmount this problem are described by Raga et al. 2013),
\item identifiable features sometimes are shifted away from a tile into neighbouring tiles in
  the image pairs (so that a shift has to
  be applied to one of the images before applying the division into tiles, see Raga et al. 2012a).
\end{itemize}

\begin{figure*}[!t]
\includegraphics[width=2.15\columnwidth]{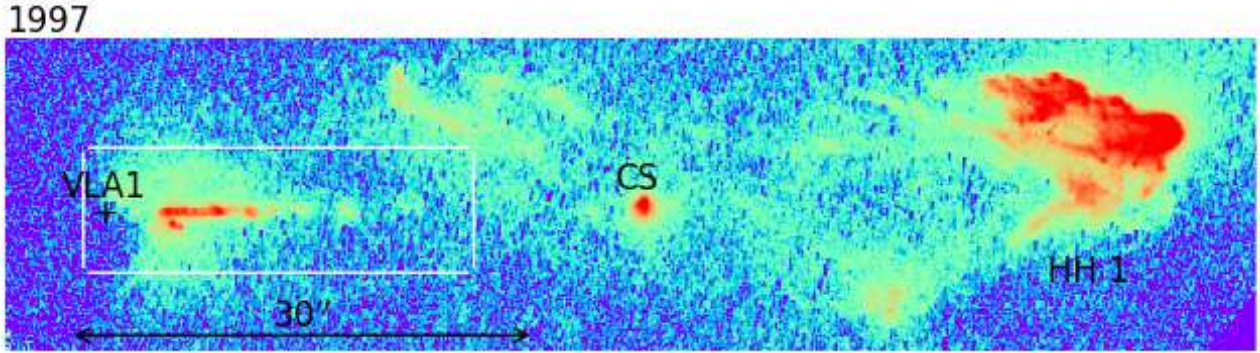}
\caption{[S~II] image taken with the HST
in 2007 of the region including the HH~1 (VLA 1) source, the HH~1 jet and
HH~1 itself. The Cohen-Schwartz (CS) star is also labeled. This image (displayed
with a logarithmic colour scale) has been rotated clockwise by $37^\circ$
so that the axis of the outflow is parallel to the abscissa. The white box encloses
the region around the HH~1 jet shown in Figures 2-4.}
\label{fig1}
\end{figure*}

\section{Measuring proper motions of HH objects with a ``wavelet technique'': a pipeline}

In order to try to avoid these problems, Raga et al. (2016a, b) proposed (and used) an
alternative, two-step method:
\begin{itemize}
\item convolving the images with a wavelet of a chosen size,
\item determining proper motions from spatial fits to the peaks in the convolved maps.
\end{itemize}
This method is of course a ``peak fitting method'', but it also incorporates a spatial
averaging (obtained through the convolution with a wavelet function) such as
is obtained with the ``cross correlation method''. The only free parameter of this
method basically is the half-width $\sigma$ of the wavelet function (and of course, the
choice of which peaks are identified as ``pairs'' in two different epochs!).

{Convolving an image with a wavelet of half-width $\sigma$ has 3 effects:
  \begin{enumerate}
  \item improving the signal-to-noise ratio at the expense of spatial resolution,
  \item eliminating emitting structures with scales $<\sigma$,
  \item eliminating structures with scales $>\sigma$.
  \end{enumerate}
  If one convolves images with functions similar to instrumental ``point spread functions''
  (e.g., with a Gaussian), one eliminates small scale structures, but larger scale structures
  in the images still remain. It is, however, unlikely that proper motions determined on
  images convolved with Gaussians would be substantially different from proper motions measured
  on convolutions with wavelets. We prefer convolutions with wavelets basically because of the
  mathematical properties of wavelet decompositions, which allow partial rebuildings of images with
  arbitrary ranges of spatial scales (see, e.g., Kajdic et al. 2012). However, this feature
  is not used in the present proper motion determinations.}

The choice of the particular form of the wavelet function does not
affect the obtained results in a substantial way.
In our implementation, we have chosen a ``Mexican hat'' wavelet:
\begin{equation}
  g_\sigma(x,y)=\frac{1}{\pi\sigma^2}\left(1-\frac{x^2+y^2}{\sigma^2}\right)\,
  e^{-(x^2+y^2)/\sigma^2}\,,
  \label{g}
\end{equation}
where $\sigma$ is the half-width of the central peak. This function has an approximately
Gaussian central peak, surrounded by a negative ring (such that its spatial integral
is zero). Together with the ``French
hat'' wavelet, this is one of the standard ``wavelet kernels''. For an
astronomically oriented discussion
of the properties of these wavelet kernels (together with graphic depictions)
see, e.g., Rauzy et al. (1993).

The convolved maps
$I_\sigma$ are then calculated through the usual integral
\begin{equation}
  I_\sigma(x,y)=\int\int I(x',y')\,g_\sigma(x-x',y-y')\,dx'dy'\,,
  \label{is}
\end{equation}
where $I(x',y')$ is the original (i.e., not convolved) image, and $(x,y)$ are
the coordinates of the convolved image. The convolutions are carried
out with a standard, ``Fast Fourier Transform'' method.

On the convolved image, we then carry out paraboloidal
fits to intensity peaks, from which we determine the positions and intensities
of the peaks. From the shifts of the positions between successive epochs, we
then determine proper motions.

We have developed a pipeline (written in Python) that:
\begin{enumerate}
\item reads an image,
\item convolves it with Mexican hat wavelets of the specified $\sigma$ values,
\item finds peaks (either chosen by the user, or searches for all peaks above
a given intensity threshold) and carries out paraboloidal fits,
\item has a ``user confirmation and labelling'' routine (with which the user
can choose the relevant peaks),
\item identifies the same peaks in two or more images and calculates the proper motions
(with linear least squares fits to the knot positions as a function of time).
\end{enumerate}

Item number 5 allows for several possibilities:
\begin{itemize}
\item the more straightforward one is to calculate
proper motions for the knots identified by the user with the same label in
the available epochs. This is of course appropriate for images with a small
number of emitting knots,
\item to automatically associate the ``nearest knots'' detected in two
successive epochs,
\item to search for the nearest knot (in the following epoch), but only
in the general direction away from the outflow source.
\end{itemize}

It is also possible to use the wavelet spectrum of the individual knots
in order to find the knot pairs that are morphologically closer to
each other, and to then use the identified pairs for calculating
proper motions. This kind of ``morphological evaluation'' using wavelet
spectra has been studied in detail by Masciadri \& Raga (2004, in
the context of search for exoplanets), but has not yet been implemented
in our pipeline.

Finally, our Python pipeline has routines to produce appropriately
labeled plots for publication. Figures 1-4 (see the
following Sections) were produced with these routines.
{After further testing and improvements to the user interface, the routine will be available
to the community.}

\begin{figure}[!t]
\includegraphics[width=\columnwidth]{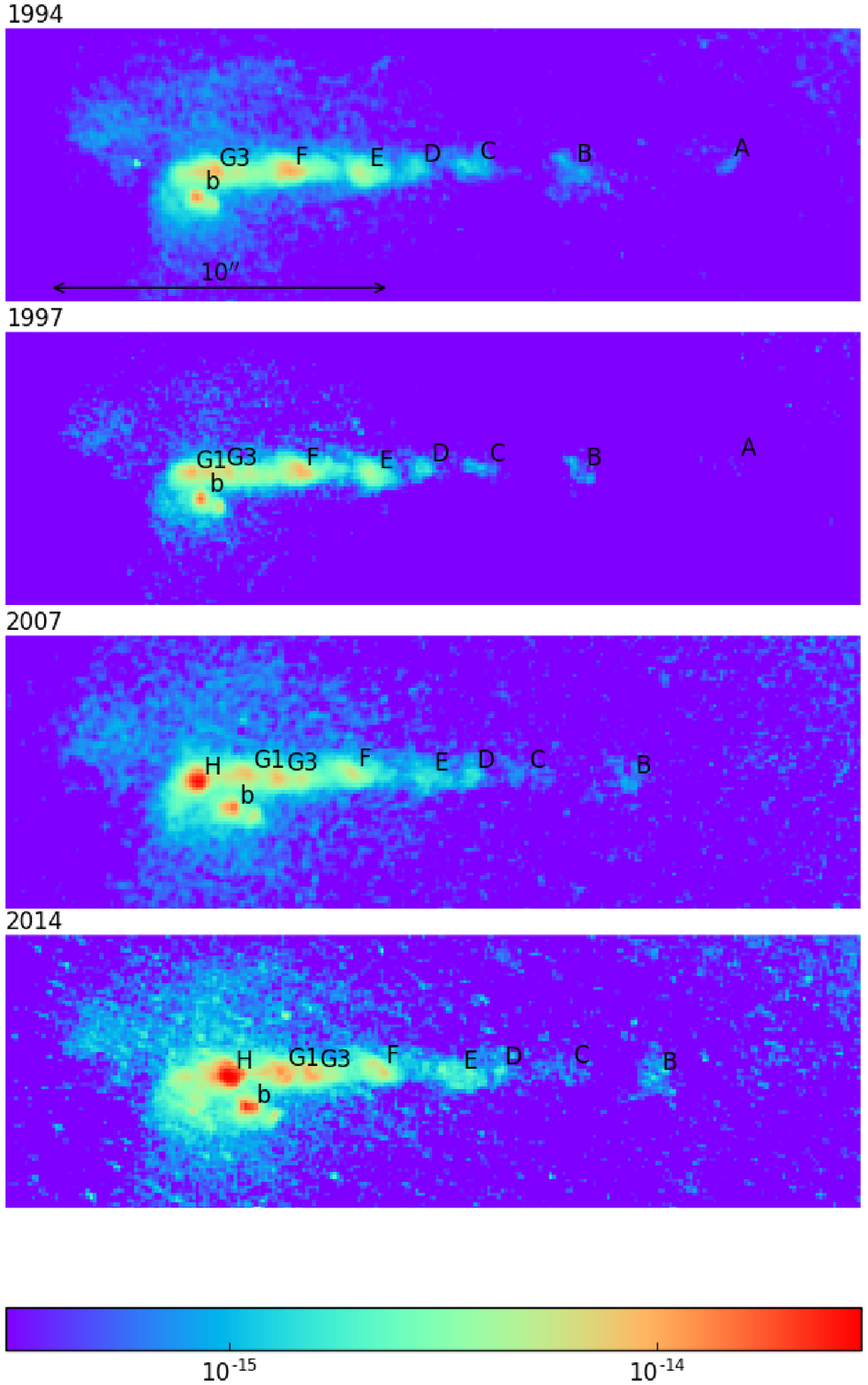}
\caption{The HH~1 jet in the four available epochs of [S~II] HST images (see Section 3). The labels
used for the knots (some of them not visible in all of the epochs) are given. The bottom bar
gives the logarithmic colour scale (in erg cm$^{-2}$ s$^{-1}$ arcsec$^{-2}$).
A flat background (of $3\times 10^{-16}$ erg cm$^{-2}$ s$^{-1}$ arcsec$^{-2}$ for the first three
epochs and of $10^{-15}$ erg cm$^{-2}$ s$^{-1}$ arcsec$^{-2}$ for the 2014 frame) has been
subtracted. The boxes have a $30''$ horizontal extent.}
\label{fig2}
\end{figure}

\begin{figure}[!t]
\includegraphics[width=\columnwidth]{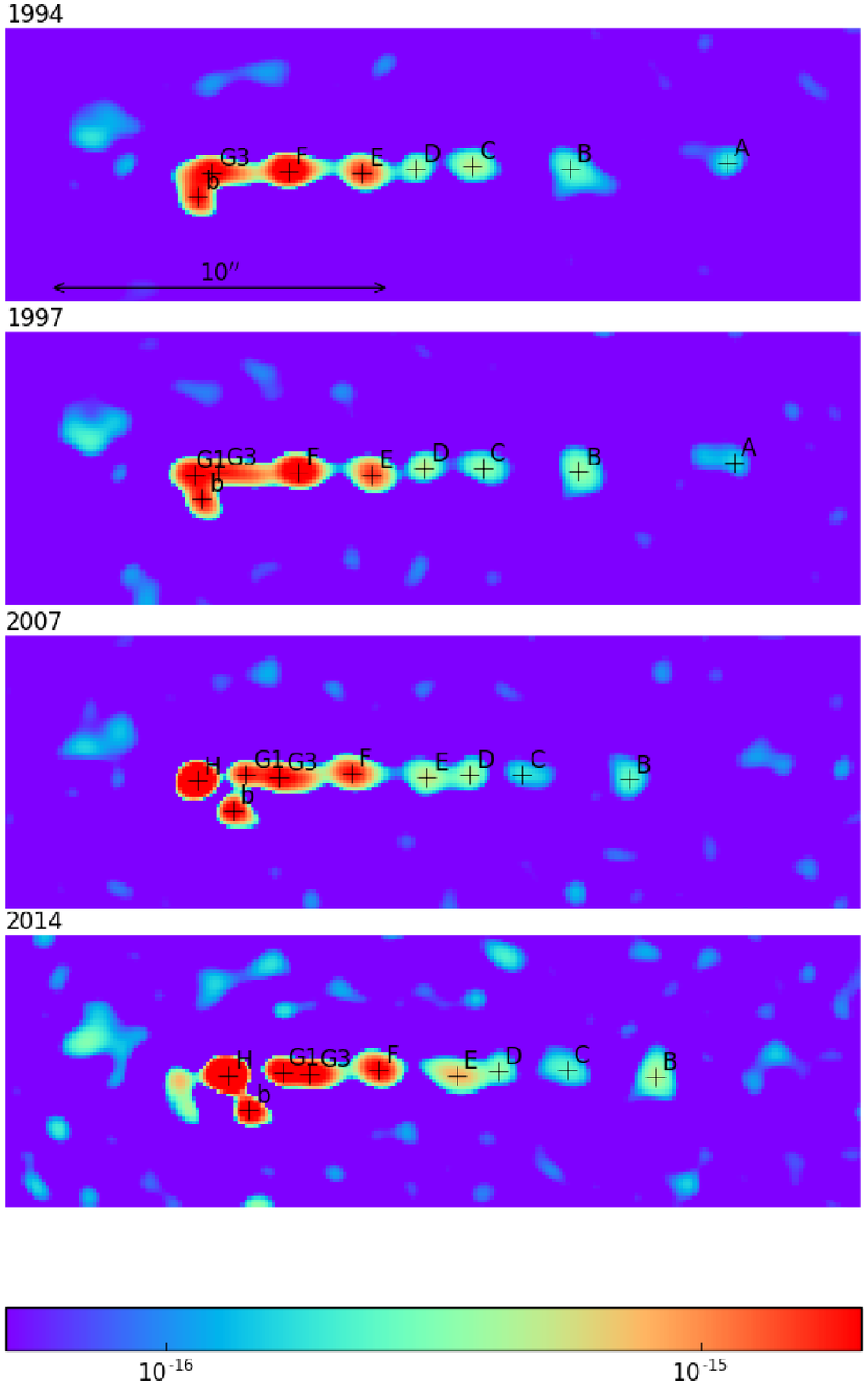}
\caption{The four epochs of [S~II] HST images (see Figure 2) convolved with a Mexican hat wavelet
of half-width $\sigma=4$~pix (see the text).}
\label{fig3}
\end{figure}

\section{Proper motions of the HH~1 jet}

We have taken the 4 epochs of [S~II] HST images of HH~1/2 described by Raga et al. (2016a, b, c)
obtained in 1994.61, 1997.58, 2007.63 and 2014.63 (we have not analyzed the H$\alpha$ frames
because the HH~1 jet is very faint in this line). Figure 1 shows a region of the 1997 frame
including the position of the VLA~1 source, HH~1 and the HH~1 jet. The Cohen-Schwartz (CS) star,
despite its strategic location, is apparently not associated with the outflow.

The analysis presented in this paper is restricted to the region around the HH~1 jet shown with
a white box in Figure 1. The [S~II] emission within this region in the four epochs is shown in Figure 2,
with the knots labeled with identifications that correspond to the ones of
Reipurth et al. (2000) and Hartigan et al. (2011). Also, we have labeled
knot B of the HH~501 jet (see Bally et al. 2002) with a lower case ``b''. This outflow appears to have been
ejected by another source in the vicinity of the HH~1/2 source (Bally et al. 2002).

In Figure 3, we show the four [S~II] frames after convolution with a $\sigma=4$~pix wavelet (i.e.,
with a central peak with a full width of $0''.8$). In these convolved frames, the jet breaks
up into knots with well defined peaks, to which we fit paraboloids (giving peak fluxes and the
positions shown in Figure 3).

With the positions measured for the successive knots (some of them seen in all frames, but others
in only two or three frames) we carry out linear least squares fits to determine their proper motion
velocities. These velocities are given in Table~1 (for a distance of 400~pc to HH~1/2) and are shown in
Figure 4.

\begin{table*}[!t]
\small
\caption{Proper motions of the HH~1 jet}
\begin{center}
\begin{tabular}{rrr}
\hline
\hline
knot & $v_x$$^a$ & $v_y$$^a$ \\
& \multicolumn{2}{c}{[km\,s$^{-1}$]} \\
\hline
A & 128 (91) & 86 (45) \\
B & 245 (12) & $-$6 (4)  \\
C & 258 (17) & 3 (4)   \\
D & 235 (12) & 5 (6) \\
E & 271 (12) & 9 (2) \\
F & 259 (12) & 18 (2) \\
G3 & 286 (13) & 9 (5) \\
G1 & 288 (16) & 17 (2) \\
H & 247 (38) & 38 (19) \\
b & 149 (12) & $-$24 (5) \\
\hline
\hline
\multicolumn{3}{l}{$^a$ the values in parenthesis} \\
\multicolumn{3}{l}{are the estimated errors} \\
\end{tabular}
\end{center}
\end{table*}

\begin{figure*}[!t]
\includegraphics[width=2\columnwidth]{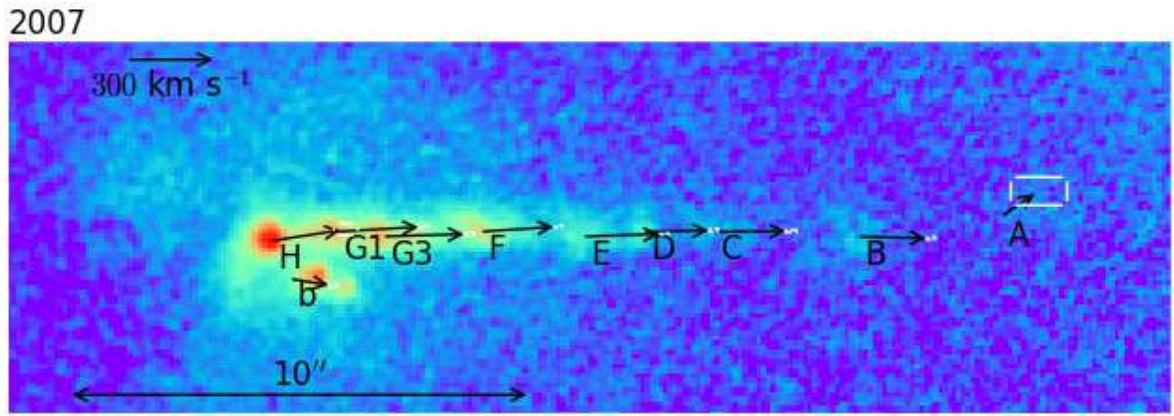}
\caption{2007 [S~II] image with the proper motions derived from the four available epochs
(see Table 1). The scale of the velocity arrows is given by the arrow in the top
left corner of the plot.}
\label{fig4}
\end{figure*}

{The errors for the proper motion velocities given in Table 1 are calculated as follows.
  We estimate that the errors of the fits to the knot positions are at most 1 pixel
  ($0''.1$). This estimated error is used to calculate the errors in the proper motion
  velocities of knots A and H, which have measured positions in only two frames. For all
  of the other knots, we calculate the error in the knot positions using the standard
  deviation of the measured positions with respect to the (straight line) least squares
  fits. These errors (as is standard) are then propagated using the covariance matrix to
  calculate the errors in the proper motion velocities. Therefore, the quoted errors
  (see Table 1) correspond to estimates of the standard deviations.}

It is clear that we do not obtain a significant result for knot A (as the errors are comparable
to the determined proper motions). This is not surprising because we only see this knot in the two
first epochs (which only have a time range of 3 years). For the remaining knots we do obtain
proper motions along ($v_x$) and across ($v_y$) the outflow axis with reasonable errors
(ranging from $\sim 10 \to 40$~km~s$^{-1}$, see Table~1).

\section{The intensities of the knots in the HH~1 jet}

Figure 5 shows the peak [S~II] intensities for knots A-I of the HH~1 jet in all
epochs, obtained through paraboloidal fits to the peaks of the convolved images (see Section 4).
It is clear that for distances from the VLA~1 source larger than $x\sim 7''$ there
is a general trend of decreasing intensities as a function of $x$. This trend
approximately follows a $I\propto x^{-3}$ power law (shown with a dashed line
in Figure 5).

Our observations do not show in a conclusive way that individual knots have
intensities that ``slide down'' the $x^{-3}$ slope as a function of time. This
is because in the 2014 frame (intensities shown with open circles in Figure 5) we
obtain systematically larger intensites for all knots than in the 2007 frame. This
is a result of the fact that the HH~1 jet region has a relatively strong reflection
nebula, with peak intensities aligned with the jet. This reflection nebula has a stronger
contribution in the 2014 frame, which was obtained with the WFC3 camera (with a
[S~II] filter of 118~\AA\ width). The first three epochs were obtained
with the WFPC2 camera (with a [S~II] filter of 47~\AA\ width), and have
a smaller contribution from the reflection continuum.

The observed $I_{[S\,II]}\propto x^{-3}$ dependence for large distances along
the HH~1 jet (see Figure 5) is in remarkable agreement with the prediction of
the analytic, ``asymptotic regime'' of periodic internal working surfaces of
Raga \& Kofman (1992). These authors note that at large enough distances from
the source, the decaying working surfaces should have an intensity $I\propto x^{-(\kappa+1)}$,
where $\kappa$ is the index of an assumed power law dependence of the line emission as
a function of shock velocity (see equation 25 of Raga \& Kofman 1992). If one takes the
plane-parallel shock models of Hartigan et al. (1987), from the lower range of the shock
velocities of their models one obtains that the [S~II] intensity has a scaling $\propto
v_{shock}^{-2}$. Therefore, the asymptotic regime of Raga \& Kofman (1992) then predicts
a [S~II] intensity $\propto x^{-3}$, in surprisingly good agreement with our observations of the
HH~1 jet (see Figure 5).

\section{The past and future evolution of the HH~1 jet}

As can be seen in Table 1, along the HH~1 jet we see a general trend of decreasing
velocities with increasing distances from the outflow source. Such a decreasing
velocity trend could in principle be the result of drag due to entrainment of
stationary, environmental material.

It is clear that in some of the ``parsec scale
HH jets'' (e.g. in HH 34, see Devine et al. 1997) a progressive decrease in proper motions
for the ``heads'' at larger distances are seen, and that this
trend cannot be explained as a result of a secularly increasing ejection velocity
from the outflow source (Cabrit \& Raga 2000). This rather dramatic slowing down
of the HH~34 ``heads'' is due to the fact that a precession of the outflow axis
results in a direct interaction of the successive heads with undisturbed environmental
material (Masciadri et al. 2002).

As the knots along the HH~1 jet are very well aligned, we would not expect them
to slow down due to frontal interaction with the surrounding, stationary environment
(as occurs in the giant HH~34 jet, see above). One might still have ``side entrainment''
into the HH~1 jet, resulting in some amount of slowing down at increasing distances
from the source. This effect has recently been evaluated by Raga (2016, in terms of
a somewhat uncertain ``$\alpha$ prescription'' for the entrainment velocity), who finds that
in order to obtain a substantial slowing down one needs a surrounding environment (in contact
with the jet beam) $\sim 10$ to 100 times denser than the jet. This is unlikely to
be the case in the optically visible HH~1 jet, which has already emerged from the
dense core surrounding the outflow source. Also (as discussed by Raga 2016), the effect
of buoyancy (which includes the gravity and the
environmental pressure gradient) is negligible for the high velocities of HH
outflows.

We therefore interpret the decreasing proper motion velocities (with increasing
distances from the outflow source) along the HH~1 jet as ballistic motions resulting
from an increasing ejection velocity as a function of time. We then
take the positions of the HH~1 jet knots in the 2007 frame, and calculate
the dynamical ejection times
\begin{equation}
t_{dyn}=-\frac{x}{v}\,,
\label{tdyn}
\end{equation}
where $x$ is the distance from the outflow source and $v$ is the proper motion velocity. In
Figure 6, we then plot $v$ as a function of $t_{dyn}$, which is the ``ballistic knot''
prediction of the past ejection time-variability history of the outflow source. The ejection
velocity has a general trend of increasing velocities towards more recent times,
which we fit with a straight line, giving:
\begin{equation}
  u_0(\tau)=(303\pm 15)+(0.59\pm 0.19) \tau\,,
  \label{vt}
\end{equation}
where $u_0(\tau)$ is the ejection velocity in km~s$^{-1}$ and $\tau$ is the ejection time in years
($\tau=0$ corresponding to 2007, since we have used the knot positions of this epoch). The linear
least squares fit (equation \ref{vt}) has been calculated with the method described in Appendix A.

The ejection velocity clearly should also have a short-term variability that produces the
knots that we observe along the HH~1 jet, so that the trend of equation (\ref{vt}) would
actually correspond to a long-term variability superimposed on the ``knot producing'' mode
(see, e.g., Raga et al. 2015c).

\begin{figure}[!t]
  \includegraphics[width=\columnwidth]{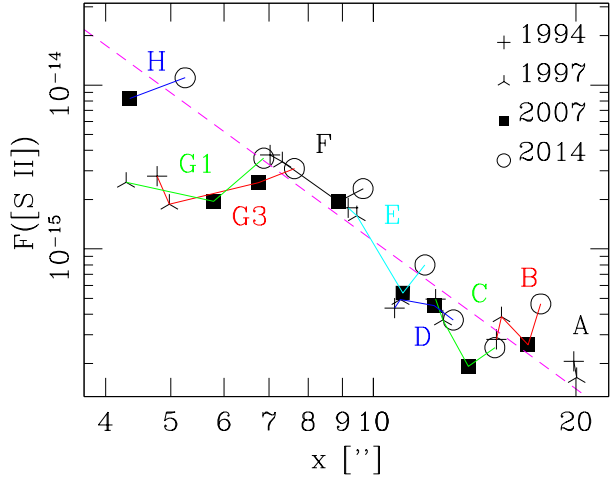}
\caption{Peak [S~II] emission of the knots (determined from the fits to the
convolved maps of the four epochs) as a function of distance from the VLA~1 outflow
source. The points corresponding to the four epochs are shown with different styles
of dots (as specified in the text above the plot) and the successive knots are
joined by lines of different colours (with labels in the same colour giving the
identifications of the knots).}
\label{fig5}
\end{figure}

\begin{figure}[!t]
\includegraphics[width=\columnwidth]{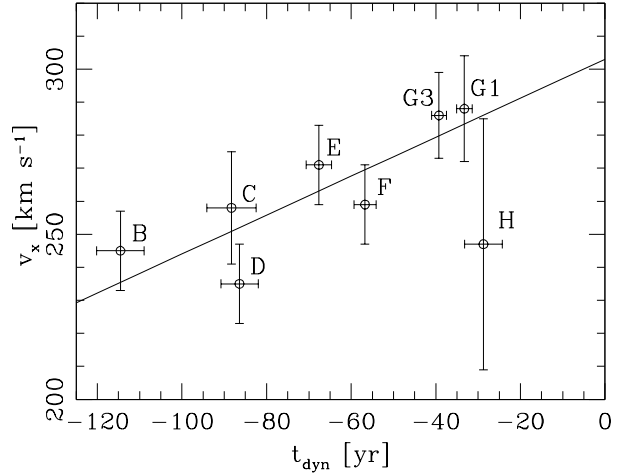}
\caption{Proper motion velocities of the knots as a function of dynamical ejection times (calculated
  through equation \ref{tdyn}, with $t_{dyn}=0$ corresponding to 2007).
  The straight line corresponds to the linear fit given in
  equation (\ref{vt}).}
\label{fig6}
\end{figure}

An interesting question is whether the relatively low velocity of knot H (see Table 1
and Figure 6) is evidence that the more recently ejected, optically detected material
is starting to show a decreasing ejection velocity vs. time trend. Given the large error of
our H knot proper motion (see Figure 6), it is hard to conclude that this is
indeed the case.

Also, we can obtain a second estimation of the motion of knot H as follows.
We take the separation between knots H and F in the
1998.15, [Fe~II]~1.64~$\mu$m image of Reipurth et al. (2000) (in which not H is
already visible), and compare it with the separation between these two knots
in our 2014.63. From this comparison, we find that knot H has an axial  motion
$17\pm 2$~km~s$^{-1}$ faster than the motion of knot F. Combining this result
with our knot F proper motion (see Table 1), we obtain a $(276\pm 18)$~km~s$^{-1}$
velocity for knot H, which is consistent (within the errors) with the
proper motion obtained from the optical images (see Table 1), but
does not support the existence of a drop in ejection velocity associated
with this knot.

{In order to use the present day positions and proper motions to predict the
future evolution of the HH~1 jet, in Figure 7 we plot the ballistic trajectories
of the knots on a $(x,t)$ plot (where $x$ is the position of the knots as
a function of time $t$). The $t=0$ axis corresponds to the 2007 knot positions.
For knot H (the trajectory with the smallest $x$ at $t=0$), we have used
the 276~km~s$^{-1}$ velocity estimated from IR images (see above).

In Figure 7, we see that the knot trajectories have crossing points in the
$x=0\to 80''$ distance range, at times smaller than $\sim 500$~yr. Therefore,
by the time that the HH~1 jet knots have reached the present-day position of
HH~1 (also shown in Figure 7), many knot-merging events will have occurred.
This result is similar to the one found by Raga et al. (2012a) for the
HH~34 jet.

\begin{figure}[!t]
\includegraphics[width=\columnwidth]{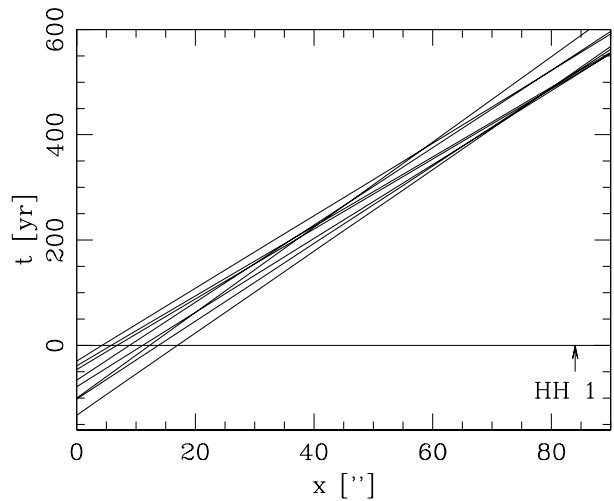}
\caption{Ballistic trajectories of the HH~1 jet knots in an $(x,t)$-plane (where $x$
  is the distance from the outflow source and $t$ is the time measured from 2007). The
  present-day position of HH~1 is indicated on the bottom right of the plot.}
\label{fig7}
\end{figure}

In order to visualize the effect of the knot-merging events, we use the
simple momentum conserving knot-merging model of Raga et al. (2012b). We
take the 2007 HH~1 jet knot positions and velocities, and assign equal masses to
all knots. We then follow the knot trajectories, merging colliding knots using
mass and momentum conservation conditions. The knots are not assigned sizes, so that
knot collisions take place at the points of trajectory crossings (see Figure 7).

The result of this exercise is shown in Figure 8. In this figure we show the knot positions
at 150~yr intervals ($t=0$ corresponding to 2007). The knots are represented as circles (centred
on the knot positions) with radii proportional to the mass of the knots. It is clear
that by $t\sim 450$~yr most of the knots have merged, and that at this time the position
of the merged knots is slightly upstream of the present position of HH~1 (at $x\approx
4.5\times 10^{17}$~cm or $75''$ at a distance of 400~pc).
Therefore, the material that is being ejected now (in the HH~1 jet) will eventually
form a new ``head'' close to the present-day position of HH~1.}

\begin{figure}[!t]
\includegraphics[width=\columnwidth]{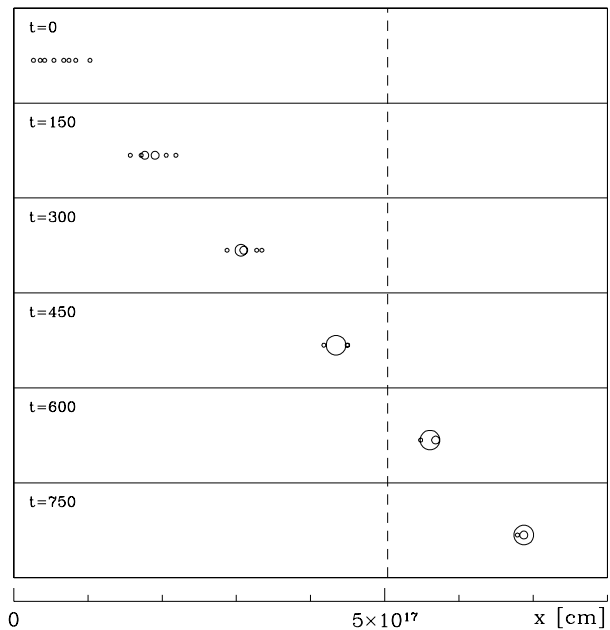}
\caption{Results of the ``momentum conserving knot'' model described in section 6. The
  $t=0$ (top) frame shows the 2007 positions of the HH~1 jet knots, and the following frames
  show the knot positions at 150~yr time intervals. The radii of the circles (indicating
  the knot positions) are proportional to the mass of the merged knots. The present-day position
  of HH~1 is shown with the dashed, vertical line.}
\label{fig8}
\end{figure}

\section{Summary}

We present a discussion of the methods that have been used in the past for measuring
proper motions of HH outflows (Section 2), and a description of a new method that
has been recently developed (Section 3). This method has been implemented in a Python
pipeline.

We then use this pipeline to determine proper motions of the knots along the HH~1 jet
in the four available epochs of [S~II] HST images (obtained in 1994, 1997, 2007 and 2014).
We find proper motions that are well aligned with the outflow axis, and with values
ranging from $\sim 230\to 290$~km~s$^{-1}$, with the faster velocities mostly in the knots
closer to the outflow source (see Section 4 and Table~1).

For each knot, we calculate a dynamical ejection time, and then plot the outflow (proper motion) velocity
as a function of ejection time (see Figure 6). This plot shows that (under the assumption of
ballistic knot motions) the outflow velocity has increased towards more recent times.

This ``acceleration'' of the ejection (as a function of ejection time) implies that the knots
presently observed along the HH~1 jet will merge into a large working surface. This is seen
in the crossings of the ballistic knot trajectories (see Figure 7) and in the momentum/mass
conserving, ``merging knot'' model shown in Figure 8. From this model, we see that
in $\sim 450$~yr most of the HH~1 jet knots will have merged, and that at this time the position
of the merged knots is slightly upstream of the present position of HH~1 (at $x\approx
4.5\times 10^{17}$~cm or $75''$ at a distance of 400~pc).
This result is qualitatively consistent with the suggestion of Gyulbudaghian (1984)
that the diverging proper motions of the condensations of HH~1 directly imply that it was formed
not far upstream from its present-day position.

This kind of morphology (a large working surface at large distances, and a short chain of
knots that will merge at the position of the present-day large working surface) is
to be expected from models of two-mode ejection variabilities (see the analytic discussion
of Raga et al. 2015c). Also, a qualitatively most similar situation has been previously
found for the HH~34 outflow (see Raga et al. 2012a, b), in which the knots along the jet will merge when
they reach the present-day position of HH~34S.

An important question is whether or not this kind of configuration (of knots along
a jet predicted to merge at the present-day position of a large ``head'') is found in other HH jets.
For some HH outflows, it is possible that proper motion data of sufficient accuracy might
be already available, and a detailed study of the available data might yield interesting
results. In other cases, future observations might be necessary in order to resolve this
question.

\acknowledgments
Support for this work was provided by NASA through grant HST-GO-13484 from the
Space Telescope Science Institute. ARa acknowledges support from the CONACyT grants
167611 and 167625 and the DGAPA-UNAM grants IA103315, IA103115, IG100516 and IN109715.
ARi acknowledges support from the AYA2014+57369-C3-2-P grant. We thank the anonymous
referee for constructive comments.

\vfill\eject
$'$
\vfill\eject
\begin{appendix}
\section{Least squares fit to the observed $v_x$ vs. $t_{dyn}$ dependence}

In Figure 6, we see that the errors in $v_x$ are probably more important than
the errors in $t_{dyn}$, so that a traditional linear, least squares fit (in which
the errors in the measured values of the abscissa are assumed to be zero) is
probably reasonable. However, in order to obtain a more convincing result, we have
done a fit in which the errors in both $v_x$ and $t_{dyn}$ are considered.

One could in principle use a standard ``errors in the two variables'' least
squares fit approach (see, e.g., the classical paper of York 1966), but
these solutions are based on the assumption that the errors in the two variables
are statistically independent from each other, and this is not the case in
our ``$v_x$ vs. $t_{dyn}$'' problem.

Given the fact that the error in the position $x$ of the knots is much
smaller than the errors in the corresponding values of $v_x$, the errors
in $t_{dyn}$ are given by:
\begin{equation}
  \epsilon(t_{dyn})=\frac{t_{dyn}}{v_x}\epsilon(v_x)\,,
  \label{et}
\end{equation}
which can be obtain by putting a perturbed velocity and appropriately
linearizing equation (\ref{tdyn}), or alternatively by using the standard
error propagation relation.

We then proceed in the standard way, writing the ``true'' values of the
measured points as $[t_{dyn}+\epsilon(t_{dyn}),\,v_x+\epsilon(v_x)]$, through
which the linear fit goes through, so that
\begin{equation}
  v_x+\epsilon(v_x)=a[t_{dyn}+\epsilon(t_{dyn})]+b\,,
  \label{vx}
\end{equation}
where $a$ and $b$ are the parameters of the linear fit that we want
to calculate. Combining equations (\ref{et}-\ref{vx}) we then find:
\begin{equation}
  \epsilon(v_x)=\frac{at_{dyn}-b-v_x}{1-\frac{at}{v_x}}\,.
  \label{evx}
\end{equation}
These are the deviations from the straight line fit resulting
from the displacements due to the errors in both $v_x$ and $t_{dyn}$.

We then define a weighted $\chi^2$ as:
\begin{equation}
  \chi^2=\sum_i \left[\epsilon(v_x)_i\right]^2w_i\,,
  \label{xi2}
\end{equation}
where the $\epsilon(v_x)_i$ values are calculated from equation
(\ref{evx}) with all of the observationally determined $(t_{dyn},v_x)$ pairs,
and $w_i=1/\sigma_i^2$, where $\sigma_i$ are the estimated errors of
the $v_x$ values (see Table 1).

Now, given the measured $(t_{dyn},v_x)$ pairs, it is straightforward
to find the values of $a$ and $b$ that give the minimum $\chi^2$ (we
do this by exploring numerically a range of values for $a$ and $b$).
It is also possible to estimate the errors of $a$ and $b$ by
perturbing (with their estimated errors) the variables of each of the measured points,
recalculating the resulting $a$ and $b$ values, and then using the standard
error propagation formula.

When this method is applied to the knots of the HH~1 jet, one obtains
$a=(0.59\pm 0.19)$~yr and $b=(303\pm 15)$~km~s$^{-1}$ (these are the values
given in equation \ref{vt}). These values are actually very similar to the
$a=(0.54\pm 0.18)$~yr and $b=(299\pm 14)$~km~s$^{-1}$ results obtained
from a standard, weighted least squares fit (in which only the errors
in the ordinate are considered).

\end{appendix}

\end{document}